\documentclass[12pt]{article}
\usepackage{epsfig,amssymb,latexsym,amsmath}
\usepackage{amssymb}  
\usepackage[cp866]{inputenc}
\usepackage[T2A,T1]{fontenc}
\usepackage[russian]{babel}

\newcommand{\be}{\begin{equation}}
\newcommand{\ee}{\end{equation}}

\usepackage{graphicx}

\title{Двухчастичное взаимодействие.
I.~Обратная задача рассеяния, нелокальные потенциалы и
концепция квантовой теории измерений}
\author{Виктор~М.~Музафаров\\
{\small Математический институт им. В.~А.~Стеклова, Моcква, Россия}\\
{\small E-mail: victor@mi.ras.ru}}
\date{}

\begin{document}

\maketitle

\begin{abstract}

Настоящая работа обобщает предшествующие работы автора~\cite{1},~\cite{2},
\cite{3}, \cite{4}, \cite{5},~\cite{6} и открывает цикл работ по общей
постановке и решению в аналитическом виде квантовомеханической обратной
задачи рассеяния (при фиксированном орбитальном моменте) в нетрадиционной
постановке: по данным рассеяния восстанавливается непосредственно волновая
функция рассеяния в импульсном представлении. Найденное решение этой задачи
задается семейством фазовоэквивалентных волновых функций с явно
контролируемым функционально-па\-ра\-мет\-ри\-чес\-ким произволом. Данный
произвол соответствует произволу выбора фазовоэквивалентных нелокальных
потенциалов и обуславливает необходимость включения в полное описание
квантовомеханических систем некоторых дополнительных предс\-тавлений
квантовой теории измерений.  Математический аппарат развиваемого подхода
иллюстрируется на примере описания нуклон-нуклонного взаимодействия в
несвязанных парциальных каналах.

\end{abstract}

\maketitle

\newtheorem{definition}{\qquad Определение}
\newtheorem{remark}{\qquad Замечание}
\newtheorem{proposition}{\qquad Утверждение}

\newcommand{\ind}{\operatorname{ind}}
\newcommand{\Var}{\operatorname{Var}}
\newcommand{\disc}{\operatorname{disc}}
\newcommand{\Ln}{\operatorname{Ln}}
\renewcommand{\ln}{\operatorname{ln}}
\renewcommand{\Im}{\operatorname{Im}}
\renewcommand{\Re}{\operatorname{Re}}

\newpage

\medskip
\hangindent=1.5cm \hangafter=0
\noindent
``Роковой порок магии заключается не в общем допущении законосообразной
последовательности событий, а в совершенно неверном представлении о природе
частных законов, которые этой последовательностью управляют. Если подвергнуть
анализу немногие примеры, то обнаружится, что они являются неправильными
применениями одного из двух фундаментальных законов мышления, а именно
ассоциации идей по сходству и ассоциации идей по смежности в пространстве и
во времени...  Сами по себе эти принципы ассоциации безупречны и абсолютно
необходимы для функционирования человеческого интеллекта. Их правильное
применение дает науку; их неправильное применение дает незаконнорожденную
сестру науки -- магию.''

\hangindent=2.0cm \hangafter=0
(James George Frazer, ``The Golden Bough'', 1st ed, Cambridge, 1890)
\bigskip

\section{Концепция сознания и квантовая теория измерений}
\label{Sec-1}

В цели и конечные задачи описания сильно-взаимодействующих двухчастичных
систем и электромагнитных и слабых процессов с их участием
физики-экспе\-ри\-ментаторы, физики-теоретики и математики явно или неявно
почти всегда вкладывают совершенно разный смысл.
\footnote{Ниже в данном пункте мы не преследуем цели концептуальной критики
каких-либо общих положений квантовой механики. Мы акцентируем внимание
только на тех положениях и их трактовке, которые будут существенно
использованы в следующих пунктах настоящей и последующих за ней работ при
выводе и интерпретации точного аналитического представления для волновой
функции сильно-взаимодействующей двухчастичной системы.}

Во-первых, это неизбежно объясняется многими причинами, связанными как с
углубляющимся расколом математики и физики как якобы самостоятельными
областями получения и/или классификации знаний, так и системой
математического и физического образования (в том числе образования,
поступающего через специализированные научные журналы и электронные
журналы), фактически углубляющей имеющийся раскол и во многих случаях
содействующей не столько приобретению новых знаний, сколько утрате уже
имеющихся знаний. (В подтверждение здесь достаточно только оценить
степень востребованности многих эпохальных работ по классической и
квантовой механике.)

Во-вторых, в значительной степени для математика описание физических
процессов заключается в построении {\it детерминированной\/} предельно
аксиоматизированной (замкнутой) системы неких образов и некоей схемы
исчисления этих образов. Дальнейшее обобщение развиваемого формализма до
некоторого обобщенного метаязыка может даже являться уже самоцелью и не
иметь никакого отношения к изначальной физической постановке задачи.
Последнее разумеется ни в малейшей степени не может служить причиной
каких-либо критических замечаний -- любая теория является объектом
собственного мира наблюдателя и вправе восприниматься наблюдателем на
равных правах с другими объектами.
\footnote{``According to the positivist philosophy of science, a physical
theory is a mathematical
model. [...] In the standard positivist approach
to the philosophy of science, physical theories live rent free in a Platonic
heaven of ideal mathematical model. That is, a model can be arbitrary detailed,
and can contain an arbitrary amount of information, without affecting the
universes they describe.'' Stephen W.~Hawking, ``G\"odel and the End of
Physics,'' talk at the Dirac Centennial Celebration, Cambridge, July, 2002.}

С точки же зрения физика многие изначальные требования математика к
построению математической модели рассматриваемых физических процессов
являются ``архитектурными излишествами''. К таковым относят зачастую,
например,

\begin{itemize}
\item[(1)] исследование классов гладкости коэффициентных функций и решений
эволюционных (динамических) уравнений -- при том (что если это необходимо
для продвижения в построении модели) предполагается, что любая подходящая
``классовая принадлежность'' имеет место,

\item[(2)] проблемы инфракрасных и ультрафиолетовых расходимостей, которые
при\-вносятся в теорию фактически руками (вследствие применения теорий вне
области их реальной применимости и апелляции к очень опасным по
разрушительности последствий для теории -- см. далее п.~2 --
абсолютизированным представлениям о ``бесконечно'' удаленных in-out
состояниях рассеяния и абсолютизированной убежденности в непогрешимости
исчисления бесконечно-малых.
\end{itemize}

В качестве комментария к п.~(1) заметим, что вследствие формального
``пренебрежения'' свойствами гладкости и асимптотического поведения
полученные аналитические решения могут не только не показывать
предполагавшейся гладкости, но и обладать разрывной сингулярной структурой.
Как правило подобные результаты представляют необычайный интерес, а
изучение структуры сингулярностей и {\it динамики\/} их развития может
вполне описывать реальную (возможно ту же) физическую систему в терминах
другого метаязыка.

Сводя вместе приведенные выше и наиболее значимые для нас предпосылки, при
последующей постановке метода обратной задачи рассеяния разделяемые нами
позиции мотивируются следующим образом.

В формировании дуальной квантовой пары \{{\it квантовое событие $\to$
квантовый объект}\} в качестве первородного нами рассматривается квантовое
событие.
\footnote{``The world is the totality of facts, not of things.''
L.~Wittgenstein, Tractatus Logico-Philo\-so\-phicus~[1922],
Routledge~\&~Keagan, London,~1955.}

До некоторой степени близкой классической аналогией такого подхода является
рассмотрение обычного языка с инверсионным порядком расположения
подлежащего (объ\-ект) и сказуемого (функция действия) -- читатель может
рассмотреть произвольное предложение, переместив сказуемое на первое место,
и убедиться, что информативность предложения при этом ощутимо меняется в
сторону описания простанственно-временной протяженности происходящего
(действия). Когда представление о потенциальном поле взаимодействия ({\it
глагол\/} действия) ставится на первое место, функция объекта (выступающего
как {\it подлежащее\/} соответствующего метаязыка теории) состоит в
констатации и переносе этого действия.

На уровне метаязыка физической теории переход от пары
$\{$~объект~$\to$~событие~$\}$ к паре $\{$~событие~$\to$~объект~$\}$
аналогичен Фурье преобразованию в классическом анализе (переходу от
конфигурационного к импульсному представлению), что несомненно должно быть
связано с формированием богатых топологических и динамических структур
метаязыка.  В частности, при дуальном рассмотрении пары $\{$~объект
$\leftrightarrow$ событие~$\}$ напрашивается введние понятия атласа
покрытий метаязыка и соответствующей дифференциальной и топологической
структур, в том числе для регулярного изучения проблем типа катастроф.
Близкое к этой идеологии изложение приложений топологических методов к
исследованию лингвистики (равно как и наоборот) можно найти в замечательном
обзоре~\cite{7}.

Детерминистское восприятие окружающего мира -- в том виде описания через
ощущения, как это присуще воспринимать человеку, т.е. через попытку свести
{\it осознание} (динамический процесс) к сознательно упрощенной
классификации событий в терминах (статических и изначально ``свободных'')
{\it объектов}, взаимодей\-ствие которых является уже чем-то вторичным, при
кажущей непротиворечивости описания физических явлений на уровне
классической механики, представляется однако именно тем источником
противоречий, которые присущи классической механике и тянут за собой шлейф
противоречий в квантовой физике.

Со времени становления Ньютоновской механики в ее окончательной форме,
представление о классической системе {\it свободных\/} материальных тел
являлось нулевым приближением описания любой динамической системы.
Взаимодействие рассматривается при этом как {\it вторичный\/} возмущающий
фактор, учет которого приводит к динамическому изменению состояния системы,
а {\it причинно-следст\-вен\-ная связь\/} различных (статических в каждый
{\it фиксированный\/} момент времени) состояний системы определяется
динамическими уравнениями движения. Однако говорить о наличии {\it
свободного\/} материального тела можно только при наличии оного, то есть
при наличии неких {\it регулярно\/} поступающих в приемные рецепторы {\it
повторяющихся\/} (воспроизводимых во времени) признаков, дающих указание на
существование данного объекта -- то есть, при наличии некоторого
дозированного (квантового, обусловленного каким-либо взаимодействием) {\it
потока воздействия\/} с данным объектом.  Далее о макроскопических объектах
как именно об объектах мы будем говорить только в указанном смысле -- то
есть, макроскопические объекты (инструмент наблюдателя) ассоциируются с
устойчивым (повторяющимся какое-то конечное время) {\it квантовым\/}
потоком {\it макроскопических\/} событий.

Регулярный поток {\it повторяющейся\/} в течение какого-либо конечного
времени информации идентифицируется в квантовом процессе {\it осознания\/}
как устойчивость некоторого события, отождествляемого сознанием с неким
объектом, обладающим определенными {\it приписываемыми ему наблюдателем\/}
свойствами. Кажущаяся субъективность данного утверждения может быть
проиллюстрирована следующим примером, встречавшимся автору в литературе. Если
взять лист бумаги, поставить на нем карандашом точку и опросить
представительную выборку наблюдателей на предмет интерпретации того, что
они наблюдают, то видимо наиболее вероятным ответом и будет ответ, что на
листе представлена точка. Но среди этой генеральной выборки вполне может
оказаться один наблюдатель (и {\it a priori\/} удалить его из генеральной
выборки никаким тестированием по Роршаху все равно не удастся), который
скажет, что на листе нарисован вид сверху на вид сбоку картины Малевича
``Черный квадрат''. То есть вопрос интерпретации наблюдений -- при условии
согласия с тем, что в процессе наблюдения каждый наблюдатель ориентируется
на свой метаязык образной интерпретации -- это самостоятельный вопрос.
Принцип конвенциональности в процессе индивидуального наблюдения
неприменим.

Дополнительно канонизируемое детерминистской Ньютоновской механикой
представление о макро-причинности (причинно-следственной связи) как о
характеристике определяемой единственно временной динамикой процессов
(событий)
\footnote{В противоположность античным представлениям о
причинно-следственной связи как для событий, так и для логических
заключений, включающей в себя {\it ценность\/} и {\it значимость\/} этих
событий и заключений.}
приводят к интерпретация некоторых устойчивых {\it событий\/} в терминах
{\it объектов}.  На наш взгляд (с точностью до замечания в последней
сноске) именно процесс перехода
$$
{\textit{квантовое событие\/}} \to {\textit{квантовый объект}}
$$
вторичный по отношению к процессу
$$
{\textit{квантовое восприятие\/}} \to
{\textit{квантовое осознание}}
$$
и определяет границы раздела квантовой и классической механик. Указанные
процессы разделены самой возможностью их проявления:  квантовое событие
есть суть рассмотрения только при его повторяемости (неединственности),
тогда как осознание (разум) выделяется именно уникальной единственностью
возникновения и проявления.  И поскольку мы имеем дело лишь с с
неединичными, повторяющимися явлениями, то набор этих явлений выступает как
{\it ансамбль}, восприятие которого разумом позволяет преобразовать
квантовый поток информации в классическую информацию о явлении с последующей
интерпретацией к понятию объекта.  Позиция автора по ряду вопросов,
связанных с уточнением этих положений, близка к работе~\cite{8}.

Квантовая механика изначально не содержит параметра, определяющего границы
ее применимости -- этот параметр является привнесенным наблюдателем и
определяется входящим в полную систему квантовой теории измерений
динамическим фактором осознания~\cite{8}. Той мерой, в какой описание
объекта в терминах указанных дискретных характеристик и непрерывных
характеристических распределений идеализируется наблюдателем уже как некое
``абсолютное'' описание объекта (в отрыве от используемого набора
метаструктур математического аппарата, анализа области применимости этого
аппарата и роли наблюдателя в процессе квантового измерения) и определяются
в частности многие трудности, присущие различным теоретическим моделям
описания квантовых систем.

Мы придерживаемся также точки зрения, что метаязык квантовой теории
должен быть открытым, так что для него может быть сформулирован объемлющий
метаязык, включающий все понятия, объекты и утверждения, но включающий
также и новые понятия, объекты и утверждения, не содержащиеся в исходном
языке. Вероятно при построении любой теории трудно удержаться от соблазна
построить именно {\it замкнутую\/} теорию, пополнив для этого при
необходимости набором опорных аксиом и опорных объектов рассмотрения
теории. Отчасти это стремление к упорядоченной классификации в рамках
определенной замкнутой теории неосознанно связано со стремлением сознания к
самосохранению в условиях нелинейного нарастания внешнего информационного
потока.

Показательный пример к последнему замечанию: где-то на рубеже 60-х и 70-х
годов применительно к научным публикациям (преимущественно в физических
журналах) наблюдалась интересная закономерность, которая на сегодня
представляет уже некое соглашение по умолчанию.  А именно, если в работе
надо было поместить графические данные, то они строились ``не по открытым
осям'' -- напротив же, сначала фиксировался некий замкнутый прямоугольник, и
только потом внутри этого замкнутого прямоугольника размещался собственно
графический материал.

На самом деле по собственному опыту нам известно, что применительно к
самому себе применение принципа строгой определенности в условиях замкнутой
системы просто опасно. Наиболее комфортные условия работы не тогда, когда
все бумаги, статьи, книги, кошки и т.д. находятся строго в определенных
местах комнаты, а, напротив, когда они довольно хаотично распределены по
самым немыслимым точкам. При том, однако, что всегда должно быть по
меньшей мере одно священное место, где вперемешку находятся только самые
нужные для работы предметы и создания (например, под диваном). В нужный же
момент сознание через разум мгновенно {\it интерпретирует\/} этот хаос
(незамкнутую систему метаобщения с окружающей средой) к некоторому
осмысленно-разумному ({\it не замкнутому\/} реально, но {\it вполне
замкнутому\/} в рамках данной {\it неформализованной\/} интерпретации)
порядку.

Следуя приведенным соображениям, в рамках развиваемого подхода к обратной
задаче рассеяния мы намеренно избегаем попыток построить замкнутую
формализованную теорию.  Если мы будем стремиться к его полной
аксиоматизации и вычленению по некоторым заложенным принципам замкнутой
системы рассматриваемых объектов, мы придем к замкнутому метаязыку имеющему
два существенных недостатка: (1)~ввести в этот метаязык любое новое
утверждение, не исключив что-то из ранее введенного, нельзя, (2)~стремление
к замыканию системы рассматриваемых объектов всегда вынуждает нас вводить
такие объекты типа ``жареного льда'' (то есть объекты, какие-то определяющие
свойства которых формально могут быть перечислены, но при этом физически
разумных аналогов этих объектов мы не знаем), которые придавая кажущуюся
завершенность конструкции привносят в нее ряд существенных неразрешимых
противоречий.

С этой точки зрения представляется полезным сначала проанализировать
имеющиеся уже в нерелятивистском случае неоднозначности
квантовомеханического описания (на примере квантовомеханического
двухчастичного одноканального рассеяния в несвязанных парциальных каналах)
и стремиться сохранить полученный функциональный произвол решения обратной
задачи рассеяния (см.~далее раздел~6).

\section{Обозначения}
\label{Sec-2}

{\bf 2.1. Конфигурационное представление.}

В предположении, что потенциал двухчастичного взаимодействия является
локальным и сферически симметричным, $v(\mathbf{r})=v(r)$,
стационарные свободная волновая функция $\psi_0(\mathbf{k};\mathbf{r})$,
волновая функция рассеяния
$\psi^{(+)}(\mathbf{k};\mathbf{r})$, $T$-матрица рассеяния
$T({\mathbf{k}}^\prime;\mathbf{k})$ и определяемая по этой амплитуде
$S$-матрица рассеяния $S({\mathbf{k}}^\prime;\mathbf{k})$
допускают, соответственно, следующие
стандартные разложения:
\footnote{Здесь и далее мы придерживаемся
обозначений и нормировок, следуя классической монографии~\cite{9}.}
\begin{align}
\psi_0(\mathbf{k};\mathbf{r})&=\Bigl(\frac{k}{\pi}\Bigr)^{\frac12}
(kr)^{-1}\sum_{l,m=0} Y^m_l(\hat {\mathbf{r}}) Y^{m*}_l(\hat {\mathbf{k}})
i^l u_l(kr),
\\
\psi^{+}(\mathbf{k};\mathbf{r})&=\Bigl(\frac{k}{\pi}\Bigr)^{\frac12}
(kr)^{-1}\sum_{l,m=0} Y^m_l(\hat {\mathbf{r}}) Y^{m*}_l(\hat {\mathbf{k}})
i^l \psi^{(+)}_l(k,r),
\\
T({\mathbf{k}}^\prime;\mathbf{k})&=\Bigl(\frac{k}{2}\Bigr)
\sum_{l,m=0} Y^m_l({\hat {\mathbf{k}}}^\prime) Y^{m*}_l({\hat {\mathbf{k}}})
t_l(k),
\\
S({\mathbf{k}}^\prime;\mathbf{k})&=
\sum_{l,m=0} Y^m_l({\hat {\mathbf{k}}}^\prime) Y^{m*}_l({\hat {\mathbf{k}}})
S_l(k).
\end{align}
Здесь ${\mathbf{k}}^\prime=k{\mathbf{r}}/r$,
сферические гармоники $Y^m_l(\hat {\,\cdot\,})$ задают собственные функции
оператора орбитального углового момента, $u_l(\,\cdot\,)$
-- функции Рикатти-Бесселя, отвечающие свободной волновой функции,
$\psi^{(+)}_l(k,r)$ -- подлежащие определению
парциальные волновые функции рассеяния,
$t_l(k^2)$ -- парциальные амплитуды $T$-матри\-цы
на массовой поверхности,
связанные представлением $t_l(k^2)=(i/\pi k)(S_l(k)-1)$
с матричными элементами $S$-матрицы рассеяния
$S_l(k)=\exp(2i\delta_l(k))$, где $\delta_l(k)$ -- функция фазового сдвига
в $l$-ом канале.

Для парциальных волновых функций рассеяния исходное стационарное
трехмерное уравнение Шредингера редуцируется к виду
\be
\biggl[-\frac{d^2}{dr^2}+\frac{l(l+1)}{r^2}-k^2\biggr]\psi^{(+)}_l(k,r)+
v(r)\psi^{(+)}_l(k,r)=0.
\ee
Из общих физических требований постановки задачи на рассеяние парциальная
волновая функция рассеяния определяется как решение данного уравнения
регулярное при $r=0$ со следующей асимптотикой на бесконечности:
\begin{align}
\psi^{(+)}_l(k,r)&\to \beta_l^{(+)}(k,r)\equiv u_l(kr)-
\frac{i}{2} (S_l(k)-1) w_l^{(+)}(kr)
\notag
\\
&\sim e^{i\delta_l(k)}\sin\Bigl(kr-\frac 12 \pi l + \delta_l(k)\Bigr)
\quad\textrm{при}\quad r\to\infty,
\end{align}
где $w_l^{(+)}(\,\cdot\,)$ -- функция Рикатти--Ганкеля.

Как для решения прямой задачи рассеяния (т.е., для
нахождения волновой функции
$\psi^{(+)}_l(k,r)$  по заданному потенциалу), так и для классической
постановки и решения обратной задачи рассеяния (т.е. нахождения функции
потенциала $v_l(r)$ по заданным данным рассеяния) чрезвычайно полезным
является введение и рассмотрение регулярного $\varphi_l(k,r)$ и нерегулярных
$f_{l,\pm}(k,r)$ решений радиального уравнения Шредингера~(5).
Эти решения выделяются, соответственно, следующими граничными условиями
\begin{align}
&\lim_{r\to 0} r^{-l-1} \varphi_l(k,r) = 1,
\\
&f_{l,\pm}(k,r) \to e^{\mp\frac 12 \pi l} w_l^{(\pm)}(kr)\qquad r\to \infty
\end{align}
и связаны между собой соотношением
\be
\varphi_l(k,r)=\frac {1}{2i}k^{-l-1}(2l+1)!!
\Bigl[ e^{-i\frac 12\pi l}\mathfrak{f}_{l,-}(k)f_{l,+}(k,r)-
e^{+i\frac 12\pi l}\mathfrak{f}_{l,+}(k)f_{l,-}(k,r)\Bigr].
\ee
Здесь $\mathfrak{f}_{l,\pm}(k)$ -- так называемые функции Йоста, задаваемые
по $S$-матрице рассеяния представлением
\be
S_l(k)=\mathfrak{f}_{l,-}(k)\mathfrak{f}_{l,+}^{-1}(k), \qquad k\in \mathbb{R}^1
\ee
и обладающие следующими свойствами
\begin{align}
\mathfrak{f}_{l,\pm}(k)&\to 1 , \qquad k\in \mathbb{R}^1, \quad |k|\to\infty,
\notag
\\
\mathfrak{f}_{l,\pm}^*(k)&=\mathfrak{f}_{l,\mp}(k), \qquad k\in\mathbb{R}^1_+,
\\
\mathfrak{f}_{l,\pm}^*(k) &\quad \textrm{аналитичны при}\quad \pm\Im k \ge 0.
\notag
\end{align}

Волновая функция рассеяния $\psi^{(+)}_l(k,r)$ выражается через регулярное
решение $\varphi_l(k,r)$ следующим образом~\cite{9}
\be
\psi_l^{(+)}(k,r)=\frac {k^{l+1}\varphi_l(k,r)}{(2l+1)!! \,\mathfrak{f}_{l,+}(k)}.
\ee

Соотношения (10)--(11) определяют краевую задачу
Римана--Гильберта~\cite{10} на пару функций $\mathfrak{f}_{l,\pm}(k)$. В
рассматриваемом в настоящей работе случае несвязанных парциальных каналов
данная задача имеет явное аналитическое решение
\be
\mathfrak{f}_{l,\pm}(k)=\exp\Bigl(-\mathcal{H}^{(\pm)}_k (\Ln S_l(k))\Bigr)=
\exp\Bigl( -2i \mathcal{H}^{(\pm)}_k (\delta_l(k))\Bigr),
\ee
где интегральный оператор $\mathcal{H}^{(\pm)}_k (\,\cdot\,)$ определяется
соотношением
\be
\mathcal{H}^{(\pm)}_k (g_l(k))=
\frac {1}{2\pi i}\int_{-\infty}^{+\infty} dq\,
\frac {g_l(q)}{q-k\mp i0}\,.
\ee

{\bf Замечание 1.}
В случае {\it связанных\/} парциальных каналов -- в том числе для наиболее
важного с физической точки зрения триплетного ${}^3S_1+{}^3D_1$ канала
$np$-рас\-сеяния, отвечающего связанному состоянию дейтрона -- решение
матричного аналога краевой задачи (10)--(11) не известно. Это отражает
общую проблему отсутствия явных решений матричных краевых задач
Римана--Гильберта -- или, в другой формулировке задач этого круга --
отсутствие явных решений систем сингулярных интегральных
уравнений~\cite{11},~\cite{12}.
\endremark

\medskip
{\bf 2.2. Импульсное представление.}

В импульсном представлении, переход к которому определяется преобразованием
\be
g_l(r) \rightarrowtail g_l(p)=
\frac {2}{\pi}\int_0^{\infty} dr\, u_l(pr) g_l(r),
\ee
радиальное уравнение Шредингера (5) принимает вид
\be
(p^2-k^2)\psi^{(+)}_l(k,p)+
\int_0^{\infty} dq\, v_l(p,q) \psi^{(+)}_l(k,q) =0.
\ee

{\bf Замечание 2.} В импульсном представлении потенциал
\be
v_l(p,q) \sim \int_0^{\infty} dr\, u_l(pr) v(r) u_l(qr)
\ee
формально ``потерял'' свою видимую локальность. На самом деле, с точки
зрения физических приложений, потенциал фактически ``потерял'' свою
локальную структуру уже в конфигурационном представлении.  Это связано с
тем, что при описании малочастичных систем и электро-слабых процессов с их
участием функция потенциала строится (то есть, восстанавливается по
двухчастичным данным рассеяния в рамках методов
Гельфанда--Левитана и Марченко, либо фитируется по свободным параметрам
закладываемого анзаца) независимо в каждой парциальной волне. То есть
функция радиального потенциала зависит от орбитального момента и следовательно
полный потенциал $v(\mathbf{r})$ нелокален.
\endremark

\medskip
Полезно ввести матричные элементы $t_l(p,k;k^2)$ так называемой
$T$-матрицы полу-вне массовой поверхности, определяемые представлением
\begin{align}
t_l(p,k;k^2)&=t_l(k,k;k^2)(p/k)^l
\notag
\\
&\quad + 2(\pi kp)^{-1}(k^2-p^2)
\int_0^{\infty} dr\, u_l(pr)
\bigl[ \psi^{(+)}_l(k,r) - \beta_l^{(+)}(k,r) \bigr]\qquad\qquad
\end{align}
и нормированные ``на массовой поверхности'' (т.е., в предельном случае
$p\to k$) соотношением $t_l(k,k;k^2)=t_l(k^2)$. Введенные матричные элементы
$t_l(p,k;k^2)$ связаны с волновой функцией в импульсном представлении
соотношением
\be
t_l(p,k;k^2)=(kp)^{-1}(k^2-p^2) \psi^{(+)}_l(k,p).
\ee

Вследствие~(12) справедливо соотношение
\be
\psi_l^{(+)}(k,p)^*=S_l^*(k)\psi_l^{(+)}(k,p),\qquad \Im k=0,
\ee
которое, будучи переписанным для полу-вне энергетической $t$-матрицы
эквивалентно условию полу-внеэнергетической унитарности
\be
\Im t_l(p,k;k^2)=-\frac 12 \pi k t_l^*(k^2) t_l(p,k;k^2),\qquad \Im k=0,
\ee

\section{Восстановление функции потенциала методами обратной задачи}
\label{Sec-3}

В традиционной постановке квантовомеханической обратной задачи предполагается,
что кинематика взаимодействия (уравнение движения) определяется уравнением
Шредингера с локальным в каждом парциальном канаде потенциалом $v_l(r)$.
Обратная задача ставится как задача восстановления потенциала $v_l(r)$ по
данным рассеяния в предположении (по умолчанию), что данные рассеяния точно
известны при всех $k\in \mathbb{R}^1_+$.

\medskip
{\bf 3.1. Метод Гельфанда--Левитана.}

В методе Гельфанда--Левитана~\cite{13},~\cite{14} данные рассеяния
изначально входят через функции Йоста, определяя ядро
\be
G_l(r,t)=\frac {2}{\pi}\int_{0}^{\infty} dk\, u_l(kr)
\biggl[ \frac {1}{\mathfrak{f}_{l,-}(k)\mathfrak{f}_{l,+}(k)}-1\biggr] u_l(kt).
\ee

Из условия полноты системы волновых функций рассеяния следует уравнение типа
Вольтерра
\be
K_l(r,t)+G_l(r,t)+\int_{0}^{r} ds\, K_l(r,s)G_l(s,t)=0,
\ee
решение которого задает явное представление для функции потенциала
\be
v_l(r)=2\,\frac {d}{dr}K_l(r,r).
\ee

\medskip
{\bf 3.2. Метод Марченко.}

В методе Марченко~\cite{15},~\cite{16} данные рассеяния входят непосредственно через данные
рассеяния $S_l(k)$, определяя ядро
\be
F_l(r,t)=-\frac{1}{2\pi}\int_{-\infty}^{+\infty} dk\,
w_l^{(+)}(kr)\bigl[S_l(k)-1\bigr]w_l^{(+)}(kt).
\ee

Из условия полноты системы волновых функций рассеяния следует уравнение типа
Вольтерра
\be
A_l(r,t)+F_l(r,t)+\int_{r}^{+\infty} ds\, A_l(r,s)F_l(s,t)=0,
\ee
решение которого задает явное представление для функции потенциала
\be
v_l(r)=-2\,\frac {d}{dr}A_l(r,r).
\ee

\medskip
{\bf 3.3. Что предпочтительнее -- конфигурационное или импульсное
представление?}

Ровно настолько, насколько с математической точки зрения важны
свойства гладкости функций исследуемых математических моделей, настолько с
физической точки зрения это может быть (и вполне оправданно) почти не
важно. В самом деле, можно верить, что любая распределенная наблюдаемая
характеристика может быть измерена (особенно в классическом случае) с любой
наперед заданной точностью. В действительности, с физической точки зрения
определение гладкости функций несколько отличается от канонического. Это
обусловлено тем, что понятие гладкости -- это понятие локальное. Из
непрерывности функции в какой-то точке не следует, что она непрерывна и в
каких-то близких точках. И математик всегда легко приведет пример функций,
непрерывных в фиксированном наборе точек, но разрывных во всех других
точках.  С точки зрения интерпретации любой физической теории в терминах
{\it наблюдаемых}, понятие разрывности привязано именно
к математической модели, поскольку свойства {\it наблюдаемых\/}
характеристик на некотором открытом множестве и на его замыкании не должны
качественно различаться -- не известно ни одного физического эксперимента,
который позволяет различить открытое и замкнутое множество.

Поэтому с физической точки зрения импульсное представление часто более
предпочтительно, чем конфигурационное представление -- в импульсном
представлении решающими являются свойства аналитичности, которые {\it не
привязаны\/} к понятию локальности. Помимо этого при изначальной постановке
и решении многих физических задач в импульсном представлении мы избегаем
опасного двойного перехода (Фурье-преобра\-зо\-ва\-ний),
на котором, в частности, и возникают проблемы неустойчивости : исходные
данные рассеяния (в импульсном представлении) $\to$ локальные
характеристики типа потенциала (в конфигурационном представлении) $\to$
наблюдаемые характеристики процессов с участием изучаемых квантовых систем
(в импульсном представлении).

\section{Выбор данных рассеяния}
\label{Sec-4}

С данного момента -- то есть, в той части, что касается конкретных
прикладных задач -- мы имеем в виду прежде всего показательное описание
общей ситуации на примере нуклон-нуклонной ($NN$) системы.

Имеющиеся на сегодня экспериментальные данные фазового анализа
$NN$-рас\-сея\-ния доступны в области $E_{\textrm{lab}}\le 3.0$~\,Gev для
$pp$-рассеяния и в области $E_{\textrm{lab}}\le 1.3$~\,Gev для
$np$-рассеяния~\cite{17}.  При этом следует различать, что
экспериментальные данные представлены как в виде серии результатов
энергетически-независя\-щего парциального анализа, так и в виде кривых
энергетически-зависящего парциального анализа, которые (естественно) не
слишком хорошо согласованы друг с другом.

Вообще говоря, приведенные в работе~\cite{17} последние данные
фазового анализа, равно как и данные предшествующих фазовых
анализов~\cite{18}--\cite{20}, нуждаются в уточнении. Не вполне
ясно, как анализировались данные по сечениям рассеяния как для
энергетически-независимого парциального анализа SP40 (когда периферическое
рассеяние должно сильно затушевывать возможности парциального анализа), так
и (в особенности) для так называемого энергетически-зависящего анализа
SP00 (насколько детально и как конкретно привлекалась модель потенциалов
OBEP-типа при параметризации амплитуд $NN$-рассеяния и как в пределах
этой подгонки распределялись 147 параметров подгонки?) Помимо этого, часть
данных, которые не согласовывались с другими данными, была отброшена при
фазовом анализе
\footnote{``In the full database, one will occasionally find experiments
which give conflicting results. Some of these have been excluded from our
fits.''~\cite{17}}
-- и это вызывает беспокойство в точности конечных
результатов.

Не входя в детали оценки точности представленных результатов
фазового анализа, мы отсылаем читателя к работам~\cite{21},~\cite{22},
где проблемы статистической обработки данных измерений обсуждаются более
детально.  Ограничимся здесь только следующими замечаниями.

В работе~\cite{17} приводится, в частности, уточнение данных по
низкоэнергетическому поведению угла смешивания $\varepsilon_1$ в триплетном
${}^3S_1+{}^3D_1$ канале при $E_{\textrm{lab}}\le 80$\,Mev.  Проводится
сравнение с данными других групп (указывается некоторое расхождение -- в
том числе вследствие того, что энергетически-независимый анализ дает две
ветви решений) и предполагается, что имеется незначительное указание на
аномально большое тензорное взаимодействие.

Вообще говоря, при предварительном анализе $np$-рассеяния в
${}^3S_1+{}^3D_1$ канале и построении волновых функций дейтрона мы еще в
1981 году (см. далее пункт~7) использовали именно данные фазового анализа
указанных альтернативных групп (Saclay). И построенные нами
более 20-ти лет назад SWF-функции
(Shirokov's Wave Functions~\cite{23}), как выясняется на сегодня,
оказываются единственными волновыми функциями, дающими согласие теории и
эксперимента для последних экспериментальных данных по поляризационному
$ed$-рассеянию (см. далее раздел~7).

Помимо этого, нам представляется очевидным, что отмечаемое
авторами~\cite{17} расхождение результатов фазового анализа
FA91~\cite{20} с последующими результатами фазового анализа ``той же
группы''
\footnote{``Somewhat larger changes are seen in comparisons with FA91.
Differences are generally largest, as one would expect, near the energy
upper limits for the various solutions and in the smaller partial
waves''~\cite{17}.}
носит абсолютно закономерный характер -- после 1992 года состав группы (см.
список авторов по ссылкам) более чем изменился и фазовый анализ следовало
производить уже и с учетом поправок на ``систематические погрешности''
участников группы.

Таким образом, предоставляемая экспериментальными данными по $NN$-рассе\-янию
информация о данных рассеяния  имеется только для узкого (конечного по
энергиям) коридора и со значительными ошибками не вполне
ясно-контроли\-руе\-мо\-го статистического характера.

В то же время хорошо известно (и, вообще говоря, изначально очевидно, в
силу того, что данные рассеяния входят в ядра уравнений Гельфанда--Левитана
и Марченко в форме Фурье-пре\-об\-ра\-зо\-ваний), что малое изменение данных
рассеяния приводит к значительным изменениям функции  потенциала. То есть
классическая постановка обратной задачи рассеяния является неустойчивой. С
учетом представленной выше ситуации по данным рассеяния
классическая постановка обратной задачи рассеяния носит скорее
академический, нежели прикладной характер. При этом в любом случае метод
Марченко носит более предпочтительный характер, поскольку в матричном
случае {\it связанных\/} каналов метод Марченко использует в качестве
входной информации непосредственно $S$-матрицу рассеяния, а не требующие
самостоятельного восстановления матрицы Йоста.

Применительно к проблеме выбора данных рассеяния,
мы исходим из того, что ``надежных'', ``независимых'',
``модельно-независимых'' и т.п.
\footnote{Подобные расплывчатые и достаточно произвольно толкуемые
описательные термины широко используются в литературе по фазовому анализу
и прикладным работам, использующим данные фазового анализа.}
данных рассеяния нет. Более того, поскольку получение таких данных связано
с опосредованными и заведомо модельными методами косвенного анализа (с
неподдающимся контролю многопараметрическим произволом), не следует
рассчитывать, что эта ситуация изменится
\footnote{``Если оценивать вероятностно-статистические методы обработки
результатов эксперимента с точки зрения физической парадигмы, то их все
придется отбросить как недостаточно надежные... с точки зрения физической
парадигмы за вероятностной моделью ошибок наблюдений вряд ли можно признать
иной статус, чем статус некоего мифа... надо быть последовательным и за
статистическими процедурами обработки информации признать статус гадания --
и это есть искусство, отдельное от самого искусства измерений.''~\cite{21}
(см.~также~\cite{22}).}.

Ничего порочного в этом мы не видим, но подобная ситуация
требует специального уточнения.
Если принять приведенные нами аргументы, исходные данные рассеяния
следует рассматривать как некие {\it распределенные\/} функции -- то есть
на общих основаниях с другими характеристиками, прямо или опосредованно
связанными непосредственно с наблюдаемыми. Опорное фазовое решение, равно
как и энергетически-зависимая ширина коридора распределения, не
детерминированы, а интерпретируются в вероятностном смысле. При этом они
должны аппроксимироваться модельным образом в рамках некоторой выбранной
динамики определенного типа, строго согласованной с выбора типа динамики
допустимых потенциальных взаимодействий.

Забегая вперед (см. далее раздел 6), пусть мы рассматриваем, например,
семейство фазовоэквивалентных волновых функций, отвечающих классу обобщенных
потенциальных взаимодействий OBEP-типа~(33). Тогда в качестве подкласса
функций $s_l(k,p)$ класса~(43) выбираются функции с областью аналитичности,
выделяемой условием
$$
s^{\textrm{OBEP}}_l(k,p) \in
\mathcal{A}^{(\infty)}(k: k\not= \pm p+i\alpha , \alpha\ge\mu, \Im k\ge 0).
$$
Соответственно, функции $h^{\textrm{OBEP}}_l(k,p)$ и семейство
$\bigl\{\psi^{(+)}_l(s^{\textrm{OBEP}}_l;k,p)\bigr\}$ будут
иметь аналитические свойства
того же типа,
$$
S^{\textrm{OBEP}}_l(k) \in
\mathcal{A}^{(\infty)}(k: k\not= i\alpha , \alpha\ge\mu, \Im k\ge 0).
$$
Наилучшим способом аппроксимации $S$-матрицы рассеяния (с воспроизведением
структуры разреза~(4) последовательностью чередующихся нулей и полюсов)
представляется хорошо развитая к настоящему времени техника диагональных
Падэ-ап\-прок\-симант.

\section{Обоснование предлагаемого метода обратной задачи}
\label{Sec-5}

Классические методы обратной задачи рассеяния (в том числе не только для
рассматриваемой в данной работе постановки -- восстановлению функции
потенциала для фиксированного орбитального момента по данным рассеяния
заданным при всех энергиях --~см., например прекрасный обзор~\cite{24})
оказываются малопригодными для физических приложений прежде всего по двум
причинам:  (1)~данные рассеяния определены слишком в узкой области энергий
и с большими ошибками, (2)~классические методы обратной задачи рассеяния
неустойчивы.

Для придания устойчивости этим методам можно апеллировать к разработке
дополняющих методов решения некорректных задач. Нам это представляется
излишним уже по одной простой причине. В основе некорректности метода обратной
задачи лежит тот факт, что восстанавливаемая функция (потенциал) не имеет
отношения к наблюдаемым характеристикам. Восстановление функций, имеющих
непосредственное отношение к наблюдаемым характеристикам --
полувнеэнергетической $t$-матрицы или волновой функции рассеяния в
импульсном представлении -- не должно давать неустойчивости. Апелляция к
функции потенциала носит скорее привычный характер, нежели оправдана
постановкой задачи. Более того, довольно странно считать, что
``распределенные'' в пространстве квантовые частицы взаимодействуют
посредством детерминированного и локального потенциала, а не посредством
столь же ``распределенного потенциального поля''. Введение же этого
распределенного потенциального поля невозможно до тех пор, пока мы изначально
привязываемся к концепции уравнения Шредингера с фиксированным (и при этом
``локальным'')
потенциалом.

Далее мы сохраним все основные требования к волновой функции рассеяния,
проистекающие из ее описания в рамках потенциального Шредингеровского
подхода, но попытаемся обобщить формализм двухчастичного
квантовомеханического описания -- в том числе с целью ввести понятие
распределенного потенциального поля и рассматривать это поле ({\it
действие}) как первичное по отношению к понятию квантовой частицы ({\it
объект}) -- см.~раздел~1 данной работы.

\section{Обратная задача рассеяния для волновой функции в импульсном
представлении}
\label{Sec-6}

Обозначим через $\mathcal{A}^{(\mu)}(k,\dots)$ класс вещественно-значных
функций $g_l(k,\dots)$ на $R^{1+\dots}$, допускающих аналитическое
продолжение по $k$ в полосу $|\Im k|\le \mu$ для некоторого
$\mu\ge 0$.

\begin{definition}
Комплексно-значная функция $S_l(k)$, заданная для всех энергий (т.е. при
всех $k\ge 0$) называется $S$-матрицей рассеяния в $l$-ом парциальном канале,
если она удовлетворяет следующим условиям:
\begin{itemize}
\item[(1)] $S_l^*(k)S_l(k)=1 \qquad \textrm{(условие унитарности)}$,
\item[(2)] $S_l^*(-k)=S_l(k)\qquad \textrm{(продолжимость
данных рассеяния на $\mathbb{R}^1_-$)}$,
\item[(3)] $S_l(k)=1+O(k^{2l+1})\quad \textrm{при}\quad k\to 0$,
\item[(4)] число связанных состояний в $l$-ом канале совпадает с индексом
$S$-матрицы, определяемым соотношением
\footnote{В данной работе мы не рассматриваем матричный случай связанных
${}^3S_1+{}^3D_1$ каналов, так что индекс $S$-матрицы равен нулю.}
\end{itemize}
\be
\ind S_l(k)\stackrel{\operatorname{def}}{=}
-(2\pi i)^{-1}\Var \Ln S_l(k)\mid^{\infty}_{0}\in \mathbb{N}_+,
\ee
\begin{itemize}
\item[(5)] $S_l(k)\in \mathcal{A}^{(\mu)}(k)$.
\footnote{В рассматриваемом случае NN-рассеяния
$\mu=m_{\pi}/2$, где $m_{\pi}$ -- масса $\pi$-мезона.}
\end{itemize}
\end{definition}

Фундаментальное значение имеет далее функция Йоста, которая безотносительно
к ее определению в рамках классической теории рассеяния формально
задается далее следующим образом
\begin{subequations}\label{e:29}
\begin{eqnarray}
\mathfrak{f}_{l,\pm}(k)&\in& \mathcal{A}^{(\mu)}(k: \pm \Im k \ge 0),
\label{1}
\\
\mathfrak{f}_{l,-}(k)&=&S_l(k) \mathfrak{f}_{l,+}(k) \quad \textrm{при}\quad \Im k =0,
\label{2}
\\
\mathfrak{f}_{l,\pm}(k)&\to& 1 \quad \textrm{при} \quad |k|\to \infty, \pm\Im k \ge 0,
\label{3}
\\
\mathfrak{f}_{l,\mp}^*(k^*)&=&\mathfrak{f}_{l,\pm}(k)\quad \textrm{при}\quad
\pm \Im k\ge 0.
\label{4}
\end{eqnarray}
\end{subequations}

{\bf Определение волновой функции рассеяния.}
{\it Для заданных данных рассеяния $S_l(k)\in \mathcal{A}^{(\mu)}(k)$ волновой функцией
рассеяния $\psi_l^{(+)}(k,r)$ будем называть комплексно-знач\-ную функцию,
удовлетворяющую следующим условиям:
\begin{subequations}\label{e:30}
\begin{eqnarray}
&&\psi_l^{(+)}(k,p)\mathfrak{f}_{l,+}(k) \quad
\textrm{-- вещественно-значная функция четности $(-1)^{l+1}$}
\notag
\\
&&\qquad\qquad\qquad\qquad\qquad\textrm{по переменным $k$, $p$},
\\
&&\langle\psi_l^{(+)}(k,p),\psi_l^{(+)}(k',p)\rangle=
2\delta (k-k'),\quad k,k'\in \mathbb{R}^1_+,\quad
\textrm{(условие ортогональности)}\qquad\qquad
\\
&&\langle\psi_l^{(+)}(k,p'),\psi_l^{(+)}(k,p)\rangle=
2\delta (p-p'),\quad p,p'\in \mathbb{R}^1_+,\quad\textrm{(условие полноты)}
\\
&&\xi_l^{(+)}(k,p)\stackrel{\textrm{def}}{=} \eta_l^{-1}(k,p)
\bigl[ \psi_l^{(+)}(k,p)-\psi_l^{(0)}(k,p) \bigr]
\in \mathcal{A}^{(\mu)}(k: \Im k\ge 0),
\\
&&h_l(k,p)\stackrel{\textrm{def}}{=}
\bigl\{\bigl[ \psi_l^{(+)}(k,p)-\psi_l^{(0)}(k,p) \bigr]
\mathfrak{f}_{l,+}(k) -2\pi^{-1}p\,\disc \mathfrak{f}_{l}(p)\eta_l(k,p)\bigr\}
\in \mathcal{A}^{(\mu)}(k,p)
\notag
\\
&&\qquad\qquad\qquad\textrm{-- вещественно-значная
функция четности $(-1)^{(l+1)}$ по $k$,~$p$}
\notag
\\
&&\qquad\qquad\qquad\textrm{и убывания $O(k^{-2})$, $O(p^{-2})$
при $|k|, |p|\to \infty$},
\end{eqnarray}
\end{subequations}
где
\begin{subequations}\label{e:31}
\begin{eqnarray}
\psi_l^{(0)}(k,p)&=&(k/p)^l[\delta(k-p)-\delta(k+p)],
\\
\eta_l(k,p)&=&a_l(k)a_l^{-1}(p),\qquad
a_l(k)=\biggl[ \frac{k}{\sqrt{k^2+\mu^2}}\biggr]^{l+1}
\\
\disc \mathfrak{f}_{l}(k)&=&\frac {1}{2i}(\mathfrak{f}_{l,+}(k)-\mathfrak{f}_{l,-}(k)(k)).
\end{eqnarray}
\end{subequations}
}

Заметим, что в данном определении волновой функции явно нигде не
предполагается, что она удовлетворяет уравнению Шредингера. Неявно же
справедливость уравнения Шредингера закладывается (как будет ясно из
последующего) в условие~(30.e).

{\bf Аналитическое представление для волновой функции рассеяния.}
{\it Ком\-плексно-значная функция $\psi_l^{(+)}(k,p)$ удовлетворяет
условиям~\thetag{30.1}, \thetag{30.4} и~\thetag{30.5}
тогда и только тогда, когда она представима в виде
\be
\psi_l^{(+)}(k,p)=\psi_l^{(0)}(k,p)+
\frac{2p\disc\mathfrak{f}_{l,+}(p)\eta_l(k,p)}
{\pi \mathfrak{f}_{l,+}(k)((k+i0)^2-p^2)}+\frac{h_l(k,p)}{\mathfrak{f}_{l,+}(k)}.
\ee
}

Доказательство данного утверждения проводится аналогично представленным
в \cite{6}, Лемма~2, выкладкам.

Представление~(32) в явной аналитической форме разделяет различные типы
сингулярности волновой функции рассеяния: (1)~первый член представления
выделяет свободную волновую функцию с $\delta$-образной сингулярностью в
импульсном представлении; (2)~второй член представления выделяет более мягкую
особенность и обуславливает точный учет так называемого ``кинематического
разреза'' -- так что при любой вещественно-значной функции $h_l(k,p)$,
определяющей третий член представления~(32), полная волновая функция
$\psi_l^{(+)}(k,p)$ удовлетворяет условию~(20); (3)~Функция $h_l(k,p)$
аналитична в полосе $|\Im k|\le m_{\pi/2}$, так что ее особенности отделены
от кинематического разреза.

В рамках $t$-матричного подхода в обобщенной
модели потенциалов однобозонного обмена
(OBEP -- One-Boson-Exchange Potentials)
особенности функции $h_l(k,p)$ определяются структурой
потенциала (см., например,~\cite{25}--~\cite{26})
$$
v_l(r)=\int_{\mu}^{+\infty} d\alpha\, \sigma(\alpha)\exp(-\alpha r)/r
$$
или
\be
v_l(p,q)=\frac{1}{2pq}\int_{\mu}^{+\infty} d\alpha\, \sigma(\alpha)
Q_l\biggl( \frac {p^2+q^2+\alpha^2}{2pq}\biggr).
\ee
Для случая, когда спектральная функция $\sigma(\alpha)$ ``насыщается''
обменом $\pi$, $\sigma$, $\eta$, $\varrho$, $\omega$ мезонами аналитическая
структура функции $h_l(k,p)$ определяется наличием семейства разрезов при
$k=\pm p +im_{\pi, \sigma, \eta, \varrho, \omega,\dots}/2$.
В энергетической $k^2$-плоскости данная структура сингулярностей
представлена на рис.~1.

\begin{figure}[ht]
\centering\epsfig{figure=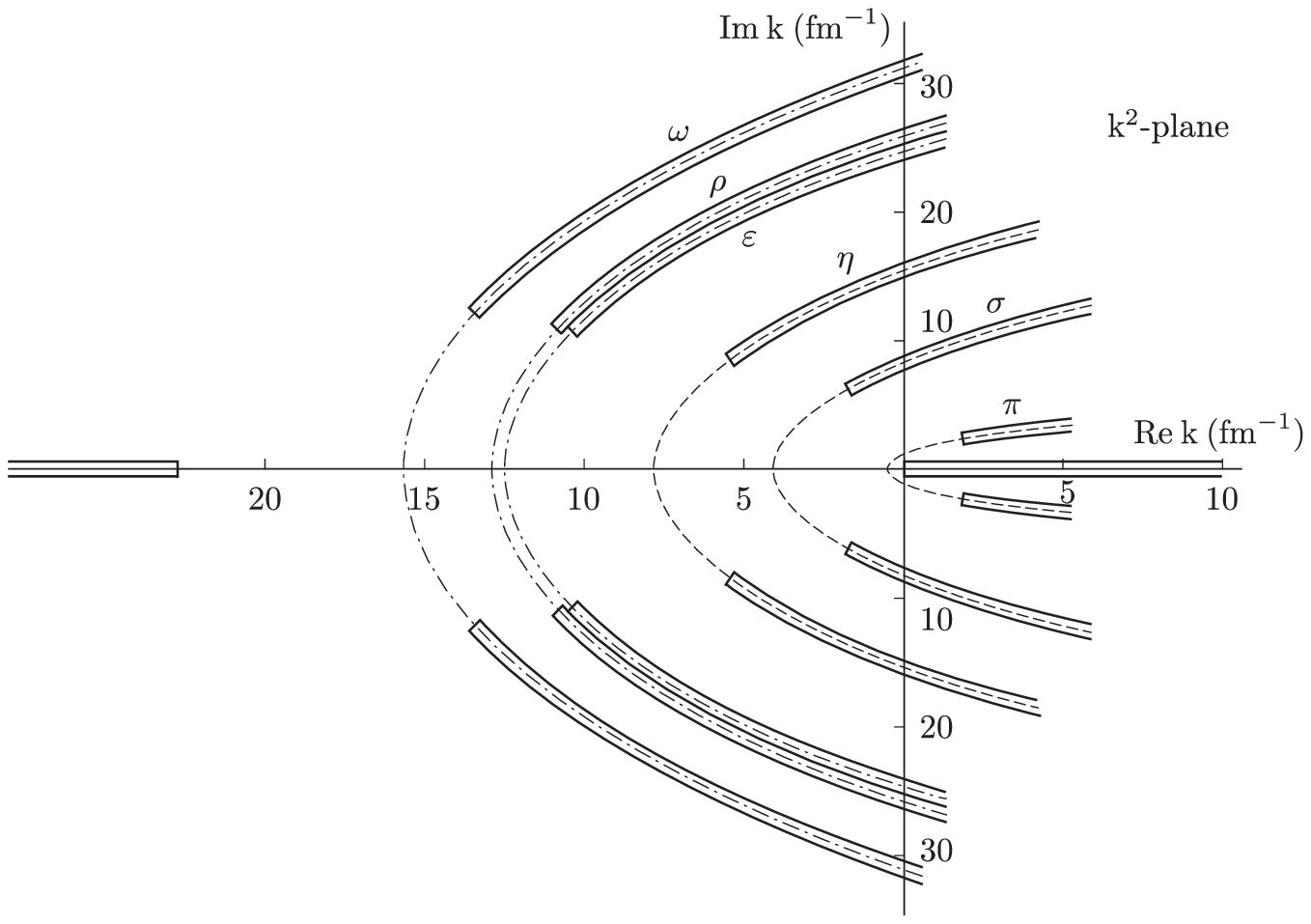,width=\linewidth}
\caption{}
\end{figure}

Не фиксируемая представлением (32) функция $h_l(k,p)$ определяется нами из
условия ортогональности~(30.b).

В стандартной потенциальной теории рассеяния справедливо следующее легко
проверяемое утверждение -- если волновая функция рассеяния удовлетворяет
уравнению Шредингера с вещественным локальным потенциалом и имеет
``правильное'' асимптотическое поведение~(6), то данная волновая функция
рассеяния автоматически удовлетворяет условию ортогональности~(30.b).  В
нашем случае справедливость уравнения Шредингера не предполагалась,
вследствие чего условие ортогональности~(30.b) является {\it
дополнительным\/} независимым условием.

Следует особо отметить, что подобный учет условия ортогональности
дает уникальную возможность введения в теорию популярных среди
разработчиков квантовых компьютеров так называемых ``запутанных''
состояний -- для этого следует лишь дополнить полный единичный оператор в
условии ортогональности ``запутывающим'' проекционным оператором
конечного дефекта.

Перейдем теперь к процедуре фиксации функции $h_l(k,p)$. Следуя~\cite{6},
мы определяем интегральный оператор
$R(\,\cdot\,,\,\cdot\,,k,p)$ равенством
\be
R(h_l,\eta_l;k,p)=\mathfrak{f}_{l,-}(p)h_l(k,p)
 -2i \mathcal{H}^{(-)}_p\bigl( \eta_l(q,p)
\disc \mathfrak{f}_{l}(p) h_l(k,p)\bigr)
\Big|_{q=p}\,.
\ee
Условие ортогональности~(30.b) переписывается для $R(h_l,\eta_l,k,p)$
как сингулярное интегральное уравнение Мусхелишвили--Омнеса вида
\be
R(h_l,\eta_l;k,p)+R(h_l,\eta_l;p,k)^*=
2\bigl[ G_l(k,p)+Q(h_l;k,p)\bigr],
\ee
где
\begin{align}
G_l(k,p)&=\pi^{-1} (k^2-p^2)^{-1}
\bigl[ k R(\stackrel{\circ}{h}_l,\eta_l;p,k)-
p R(\stackrel{\circ}{h}_l,\eta_l;k,p)\bigr],
\\
\stackrel{\circ}{h}_l (k,p)&=\eta_l(k,p)\disc \mathfrak{f}_{l}(p),
\\
Q(h_l;k,p)&=-\frac 14 \langle h_l(k,q), h_l(p,q)\rangle_{q}\,.
\end{align}

Данное сингулярное интегральное уравнение имеет аналитическое решение
\be
h_l(k,p)=\overline{\mathcal{L}}_k^{-1}(a_l;\overline{h}_l(k,p)),
\ee
где функция $\overline{h}_l(k,p)$ удовлетворяет регулярному интегральному
уравнению
\be
\overline{h}_l(k,p)=\overline{\mathcal{L}}_k^{+1}(a_l;\,\cdot\,)
\overline{\mathcal{L}}_p^{+1}(a_l;\,\cdot\,)
\bigl( G_l(k,p)+s_l(k,p)\bigr)
-\frac 14 \langle \overline{h}_l(k,q),\overline{h}_l(p,q)\rangle_q.
\ee

Здесь на пробной функции $g_l(k)$ интегральные операторы
$\overline{\mathcal{L}}_k^{\pm 1}(a_l;\,\cdot\,)$ задаются по
определяемым в~(14) операторам $\mathcal{H}^{(\nu)}_k (\,\cdot\,)$
следующим образом:
\begin{gather}
\overline{\mathcal{L}}_k^{\pm 1}(a_l;g_l(k))=
\sum_{\nu=\pm} \nu
\mathcal{H}^{(\nu)}_k (\mathfrak{f}_{l,\pm}(\nu k)\mu_{l,\pm}(k) g_l(k)),
\\
\mu_{l,+}(k)=\bigl(\mathfrak{f}_{l,+}(k) \mathfrak{f}_{l,-}(k)\bigr),
\qquad \mu_{l,-}(k)\equiv 1,
\end{gather}
а вещественно-значная функция $s_l(k,p)$ ограничена единственно условиями
\begin{align}
s_l(k,p) &\in \mathcal{A}^{(\mu)}(k: \Im k\ge 0),
\\
s_l(k,p) &\textrm{-- антисимметричная по переменным $k,p$
вещественно-значная функция}
\notag
\\
&\textrm{четности $(-1)^{(l+1)}$ по этим аргументам
и убывания $O(k^{-2})$, $O(p^{-2})$}
\notag
\\
&\textrm{при $|k|$, $|p|\to \infty$}.
\end{align}

Заметим, что все члены уравнения (40), за исключением функции $s_l(k,p)$,
отвечают функциям симметричным по переменным $k,p$. Так что именно остающаяся
незафиксированой функция $s_l(k,p)$ отвечает выбору конкретного
фазовоэквива\-лентного потенциала~(см.~\cite{6}).

Решение уравнения~(40) существенно
упрощается за счет того, что семейство функций $s_l(k,p)$, параметризующих
класс фазовоэквивалентных волновых
функций, образует ``звездное'' семейство.
А именно, справедливо следующее

\begin{proposition}
Пусть функции $h_l(s^{(1)}_l;k,p)$, $h_l(s^{(2)}_l;k,p)$ и
$h_l(s^{(\alpha)}_l;k,p)$ отвечают решениям
уравнения~\thetag{40}, параметризованным функциями
$s^{(1)}_l(k,p)$, $s^{(2)}_l(k,p)$ и
$s^{(\alpha)}_l(k,p)=\alpha s^{(1)}_l(k,p)+ (1-\alpha) s^{(2)}_l(k,p)$,
соответственно, при произвольном $\alpha \in (0,1)$. Тогда справедливо
тождество
\be
h_l(\alpha s^{(1)}_l+ (1-\alpha) s^{(2)}_l;k,p) \equiv
\alpha h_l(s^{(1)}_l;k,p) + (1-\alpha)h_l(s^{(2)}_l;k,p),
\ee
задающее на семействе функций $s_l(k,p)$, параметризующих семейство
решений основного уравнения~\thetag{40}, гомотопию с $\alpha \in (0,1)$.
\end{proposition}

Доказательство этого утверждения основывается на том, что гомотопия с
$\alpha \in (0,1)$ не затрагивает первые два члена  основного
представления~(32) (т.е., оставляет инвариантным суммарный вклад свободной
волновой функции и кинематического разреза. Так что, вообще говоря,
уравнение~(40) достаточно решить только в одном частном случае
$s_l^{{0}}(k,p)\equiv 0$.

Вследствие гомотопии (44) на пространстве волновых функций возникает
структура расслоения, которая нуждается в отдельном изучении.

\section{Некоторые практические результаты}
\label{Sec-7}

В простейшем варианте представленные выше результаты использовались нами
ранее~\cite{23} для описания $NN$-рас\-сеяния в триплетном ${}^3S_1+{}^3D_1$
канале, отвечающем связанному состоянию дейтрона. При этом -- в отсутствие
имеющегося на сегодня понимания проблемы -- вклад нефизических разрезов (то
есть, учет функции $h_l^(k,p)$ имитировался многополюсной аппроксимацией
разрезов (см. рис.~1), положение и вычеты в которых фитировались из условия
``минимизации функции дефекта условия унитарности''. Данные рассеяния в
области $E_{\textrm{lab}}\le 0.3$~\,Gev выбирались следуя фазовому анализу
Saclay (так что, в частности, функция $\varepsilon_1(k)$  меняла знак при малых
энергиях см.~\cite{23}.) При б\'ольших энергиях для определения функций
фазовых сдвигов мы (следуя~\cite{25},~\cite{26}) использовали спиральные
амплитуды Regge-анализа и в асимптотической области привлекали кварковую
модель~\cite{27}.  Требуемая для построения волновой функции дейтрона
матрица Йоста строилась  по специальной теории возмущений~\cite{28}.

Заметим, что если для случая несвязанных парциальных каналов функция Йоста
тождественно совпадает с определяемой интегральным представлением~(13)
функцией, то в матричном случае это не так. И матричная функция Йоста в
стандартной потенциальной теории имеет особенность типа $k^{-2}$ в одном из
каналов. Мы сознательно игнорировали в~\cite{28} эту особенность,
полагая, что потенциальную теорию рассеяния следует использовать только в
области ее применимости.

Любопытно, что и сейчас построенная нами в 1981 году волновая функция
дейтрона (SWF -- Shirokov's Wave Function)
\footnote{В некоторых работах -- см.,~например,~\cite{29} и приведенные там
более ранние ссылки -- данная волновая функция именуется MT-волновая
функция (волновая функция Музафарова--Троицкого). Подобное поименование
следует вероятно публикации~\cite{30}, в которой эта волновая функция
использовалась для описания процесса упругого $ed$-рассеяния и вводилось
поименование MT-approach со ссылкой на~\cite{31}.  Автор настоящей
публикации не несет ответственности за оба поименования и именует
построенные в~\cite{23}, стр.~90, и численно
протабулированные позже в~\cite{31} волновые
функции по их правовой принадлежности как Shirokov's Wave Functions --
см. работу~\cite{32}, в которой видимо впервые было получено представление,
близкое к~(32).}
блестяще описывает новые экспериментальные данные по упругому
$ed$-рассеянию. Более того, полученные недавно экспериментальные данные по
поляризационному $ed$-рассеянию описываются только данной волновой функцией
(см. рис.~2).

\begin{figure}[ht]
\centering\epsfig{figure=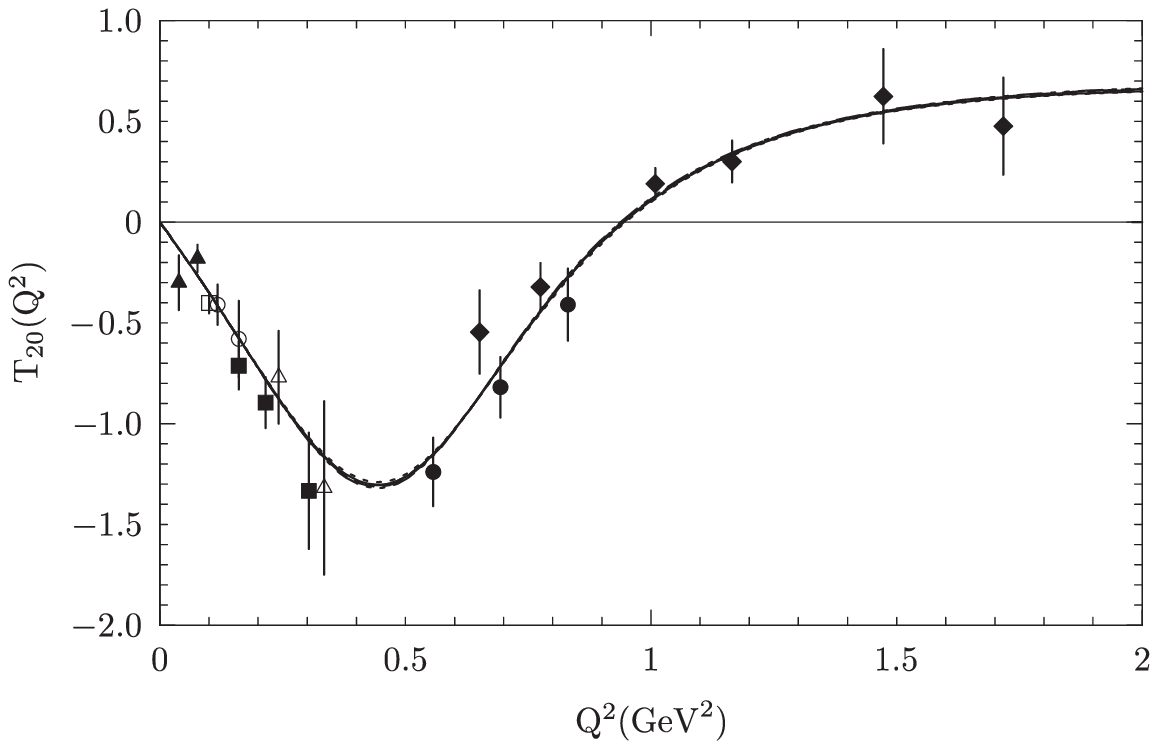,width=\linewidth}
\caption{}.
\end{figure}

На наш взгляд, это обусловлено тем, что данные волновые функции мягко
учитывают тензорное взаимодействие и не привязаны чрезмерно строго к
потенциальной теории мезонных обменов. Последнее прекрасно иллюстрируется,
в частности, тем, что эти функции
плохо фитируются стандартной подгонкой волновых функций под функции типа
Парижского потенциала -- см., например, ошибочную публикацию~\cite{33}, где
коэффициенты Тэйлоровского разложения (являвшиеся коэффициентами подгонки
Shirokov's Wave Function) нарастают на первых восьми членах
разложения по $e^{-m_\pi r}$ с единицы до $10^7$!

В следующей работе мы рассмотрим ослабленную (аксиоматически
не замкнутую) формулировку квантовой механики, в рамках которой описание
квантовомеханических систем проводится без привлечения представлений об
асимптотически свободных состояниях рассеяния.

Многие вопросы, связанные прежде всего с результатами представленными в
пункте~6, обсуждались автором с профессором Роджером Ньютоном во время его
визита в ЛТФ ОИЯИ (Дубна). И автор признателен профессору Роджеру Ньютону
за многочисленные стимулирующие замечания.

\end{document}